\documentclass[prb,twocolumn,aps]{revtex4}
\usepackage{graphicx}
\newcommand{\beq}{\begin{eqnarray}}
\newcommand{\eeq}{\end{eqnarray}}
\newcommand{\bs}{\bar\sigma}

\newcommand{\tn}{\tilde n}
\begin{document}
\title{Breakdown of Fermi Liquid Theory
in Doped Mott Insulators by Dynamical Spectral Weight Transfer}
\author{Philip Phillips, Ting-Pong Choy, and Robert G. Leigh}

\affiliation{Department of Physics,
University of Illinois
1110 W. Green Street, Urbana IL 61801, U.S.A.}

\begin{abstract}
We show that doped Mott insulators exhibit a collective degree of
freedom, not made out of the elemental excitations, because the number
of single-particle addition states at low energy per electron per spin
is greater than one. The presence of such a collective degree of
freedom which is not a consequence of proximity to a phase transition
but rather stems from dynamical spectral weight transfer from high to low energies. This physics is captured by the charge $2e$ boson
that emerges by explicitly integrating out the high-energy scale in
the Hubbard model. The charge $2e$ boson binds to a hole, thereby
mediating new charge $e$ states at low energy. It is the presence of
such charge $e$ states which have no counterpart in the non-interacting
system that provides the general mechanism for the breakdown of Fermi
liquid theory in doped Mott insulators. The relationship between
the charge $2e$ boson formulation and the standard perturbative treatment is explained.
\end{abstract}

\maketitle

The cornerstone of the standard theory of metals, Fermi
liquid theory, is that although the electrons interact, the
low-energy excitation spectrum stands in a one-to-one
correspondence with that of a non-interacting system. In the normal
state of the copper-oxide high-temperature superconductors, drastic
deviations from the Fermi liquid picture obtain. The most dramatic
effects include the 1) presence of a pseudogap\cite{alloul,norman,timusk} in the single-particle
spectrum, 2) broad spectral features in the electron-removal spectrum\cite{removal}, 
and 3) the ubiquitous $T-$ linear resistivity\cite{batlogg} above the
superconducting transition. In the electronic state that accounts for this physics, the key assumption of Fermi
liquid theory that the low-energy spectra of the interacting and free
systems bare a one-to-one correspondence must break down. More precisely, the
interacting system must contain electronic states at low energy that
cannot be derived from the non-interacting system.  
As Polchinski\cite{polchinski} and others\cite{shankar} have shown
breaking Fermi liquid theory is notoriously difficult.  Explicitly, aside from
Cooper pairing, all 4-fermion interactions are
irrelevant\cite{polchinski} once one posits that electrons are the
charge carriers. Hence, some fundamentally new
degree of freedom is needed.   How might such new degrees of freedom arise?  One
possibility\cite{ruckenstein} is if spectral weight transfer between high and low
energies mediates new electronic states at low energy. Such states
will have no counterpart in the non-interacting system, thereby
leading to a breakdown of Fermi liquid theory.  In this paper, we show
explicitly that this state of affairs obtains quite generally in a
doped Mott insulator. The new electronic state at low energies
correspond to a bound state between a doubly occupied site and a hole.
This new electronic state is mediated by the collective charge $2e$ boson
which emerges\cite{lowen1} in the exact low-energy theory of a doped Mott
insulator described by the Hubbard model. Our work stands in contrast to a recent continuum non-Fermi liquid proposal\cite{leclair1,leclair2} whose relationship to lattice Mott physics is not established.

\begin{figure}
\centering
\includegraphics[width=8.0cm,]{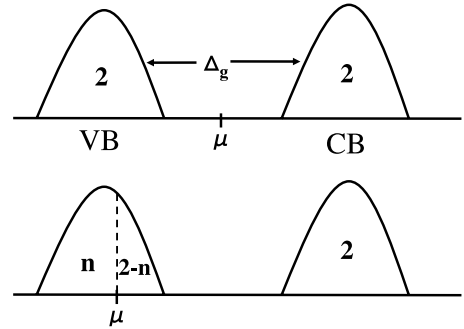}
\caption{Evolution of the single-particle density of states in a Fermi
  liquid in the valence band as a function of the electron filling, $n$.  The total weight of the valence band is a constant 2, that
  is, 2 states per site.  Doping simply pushes one state above the
  chemical potential.  The integral of the density of states below the
chemical potential is always the filling, $n$.}
\label{fig1}
\end{figure}

\section{State Counting in Fermi Liquids}

\begin{figure}
\centering
\includegraphics[width=9.0cm]{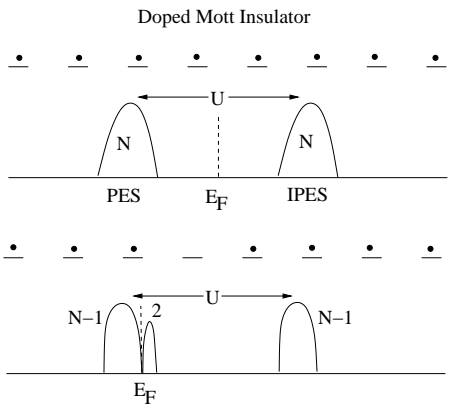}
\caption{Evolution of the single-particle density of states from
  half-filling to the one-hole limit in a doped Mott insulator
  described by the Hubbard model.  Removal of an electron results
  in two empty states at low energy as opposed to one in the
  band-insulator limit. The key difference with the Fermi liquid is
  that the total weight spectral weight carried by the lower Hubbard
  band (analogue of the valence band in a Fermi liquid) is not a
  constant but a function of the filling. }
\label{fig2}
\end{figure}

In a non-interacting system, the number of low-energy addition states
per electron per spin is equal to one. Should the number of low-energy
addition states per electron per spin exceed unity, Fermi liquid
theory fails and new electronic states emerge at low energy that
cannot be constructed from the non-interacting system. To show that
this state of affairs obtains in a doped Mott insulator, we compare
the number of electrons per site ($n_h$) that can be added to the
holes created by the dopants with the number of single-particle
 addition states per site at low energy,
\beq\label{dos}
L=\int_\mu^\Lambda N(\omega)d\omega,
\eeq
defined as the integral of the single-particle density of states
($N(\omega)$) from the chemical potential, $\mu$, to a cutoff energy
scale, $\Lambda$, demarcating the IR and UV scales.  
Consider first
the case of a Fermi liquid or non-interacting system.  As illustrated
in Fig. (\ref{fig1}), the total weight of the valence band is 2, that
is, there are 2 states per site.  The integrated weight of the valence
band up to the chemical potential determines the filling.
Consequently, the unnocupied part of the spectrum, which determines $L$,
is given by $L=2-n$. The number of electrons that can be added to the
empty sites is also $n_h=2-n$ (see Fig. (\ref{fig2})).  Consequently, the number
of low-energy states per electron per spin is identically unity. The
key fact on which this result hinges is that the total weight of the
valence band is a constant independent of the electron density. 

\section{Doped Mott Insulators: Not just electrons}

For a doped Mott insulator, the situation is quite different as a
charge gap splits the spectrum into lower and upper Hubbard bands (LHB
and UHB, hereafter)
depicted in Fig. (\ref{fig2}).  At half-filling the chemical potential
lies in the gap.  The sum rule that 2 states exist per site 
applies only to the combined weight of both bands.  At any finite
doping,
the weight in the LHB and UHB is determined by the density.    For
example, at half-filling each carries half the spectral weight. Even
in the atomic limit, the spectral weights in the LHB and UHB are
density dependent as shown in Fig. (\ref{fig2}).
Nonetheless, for a doped Mott insulator in the atomic limit, i.e. one
electron per site with infinite on-site repulsion $U$, it is still true that $L/n_h=1$, because creating a hole leaves behind an empty site which
can be occupied by either a spin-up or a spin-down electron. Hence,
when $x$ electrons (see Fig. (\ref{dos})) are removed, $L=n_h=2x=2-2n$
in the atomic limit. Recall for a Mott system, $x=1-n$ as the hole
doping occurs relative to half-filling.  Hence, for a Mott system in
the atomic limit, the total weight in the LHB increases from 1/2 at
half-filling to $1-x+2x=2-n$ as the system is doped.  This result
illustrates that the total weight in the LHB goes over smoothly to the
non-interacting limit when $n=0$.  That is, 2 states exist per site at
low energy entirely in the LHB.  The change from half the spectral
weight at $n=1$ to all the spectral weight residing in the LHB at
$n=0$ is a consequence of spectral weight transfer.  The atomic limit,
however, only captures the static (state counting) part of the
spectral weight.  In fact the $2x$ sum rule,in which $L/n_h=1$, is
captured by the widely used\cite{lee,lee1,lee2} $t-J$ model of a doped
Mott insulator in which no doubly occupied sites are allowed.   However, real Mott systems are not in the
atomic limit. Finite hopping with matrix element $t$ creates double
occupancy, and as a result empty sites with weight $t/U$. Such empty sites with fractional weight contribute to $L$. Consequently,
when $0<t/U\ll 1$, $L$ is strictly larger than $2x$. The actual behaviour,
red curve in Fig. (\ref{dos}), has been confirmed
both experimentally by oxygen 1s x-ray-absorption spectroscopy\cite{xray} and
theoretically by exact diagonalization\cite{diag} and
perturbation theory\cite{harris}. The number of {\it electrons}\cite{frac} that can be
added at {\it  low energy} remains twice the number of dopants, $n_h=2x$, even
when the hopping
is finite. That is, the quantum fluctuations mediated by the hopping
do not change the number of electrons that can be added at low energy.
As a result, for a doped Mott insulator,
$L/n_h>1$. Consequently, in contrast to a Fermi liquid, simply counting
the number of electrons that can be added does {\it not} exhaust the
available phase space to add an electron at low energy. Thus,
additional degrees of freedom at low-energy, not made out of the elemental
excitations, must exist. They arise from the hybridization with the doubly occupied sector
and hence must emerge at low energy from a collective charge 2e
excitation. The new charge $e$ state that emerges at low energy must correspond to a
bound state of the collective charge $2e$ excitation and the hole that
is left behind. It is the physics of this new charge $e$ state that
mediates the non-Fermi liquid behaviour in a doped Mott insulator.
\begin{figure}
\centering
\includegraphics[width=3.0cm,angle=-90]{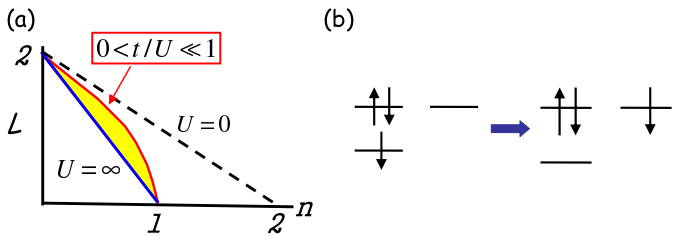}
\caption{a) Integrated low-energy spectral
weight, $L$, defined in Eq. (\ref{dos}), as a function of the
electron filling, n: 1) the dashed line is the non-interacting limit, vanishing on-site
interaction ($U=0$), in which $L=2-n$, 2) atomic limit
(blue line) of a doped
Mott insulator, $U=\infty$, in which $L=2(1-n)=2x$, $x$ the doping
level and 3) a real Mott insulator in which $0<t/U\ll 1$, red curve. For
$0<t/U\ll 1$, $L$ must lie strictly above the $U=\infty$ limit and hence $L>2x$ away from the
atomic limit. (b) Hopping processes mediated by the $t/U$ terms in
the expansion of the projected transformed operators in terms of the bare electron operators (see Eq. (\ref{trans})). As a
result of the $t/U$ terms in Eq. (\ref{eq:sc1}), the low-energy
theory in terms of the bare fermions does not preserve double
occupancy. The process
shown here illustrates that mixing between the high and low-energy
scales obtains only if double occupancy neighbours a hole. In
the exact low-energy theory, such processes are mediated by the new
degree of freedom, $\varphi_i$, the charge $2e$ bosonic field which
binds a hole and produces a new charge $e$ excitation, the collective
excitation in a doped Mott insulator. }
\label{dos}
\end{figure}

In a series\cite{lowen1} of recent papers, we found the collective charge mode in a doped Mott insulator by integrating out exactly
the high-energy scale in the Hubbard model, thereby obtaining an exact description of the IR physics. Consistent with the physical argument presented above, the collective mode is a
charge $2e$ bosonic field which mediates new electron dynamics at low
energy and is not made out of the elemental excitations in the UV. In
an attempt to clarify this theory, we establish the relationship
between the standard perturbative approach and the physics mediated by the charge $2e$ boson. By comparing how the operators
transform in both theories, we are able to unambiguously associate
the charge $2e$ boson with dynamical (hopping-dependent) spectral weight
transfer across the Mott gap\cite{diag,eskes,slavery}. Further, the
spectral weight transfer is mediated by a bound excitation of the
charge $2e$ boson and a hole in accord with the physical argument
presented above.

\section{Standard Approach}

Of course the standard approach for treating the fact that $L/n_h>1$
in a doped Mott insulator is through perturbation theory, not by explicitly constructing the missing degree of freedom. The goal of the perturbative approach is to bring the Hubbard model
\beq
H&=&-t\sum_{i,j,\sigma} g_{ij}
a^\dagger_{i,\sigma}a_{j,\sigma}+
U\sum_i a^\dagger_{i,\uparrow}a^\dagger_{i,\downarrow}a_{i,\downarrow}a_{i,\uparrow}
\eeq
into block diagonal form in which each block has a fixed number of `fictive' doubly occupied sites. Here $i,j$ label lattice sites,
$g_{ij}$ is equal to one iff $i,j$ are nearest neighbours and $a_{i,\sigma}$
annihilates an electron with spin $\sigma$ on lattice site $i$. We say `fictive' because the operators which make
double occupancy a conserved quantity are not the physical electrons but
rather a transformed (dressed) fermion we call $c_{i\sigma}$
defined below.
Following Eskes et al.\cite{eskes},
for any operator $O$, we define $\tilde O$ such that
$ O\equiv {\bf O}(a)$ and $\tilde{O}\equiv {\bf O}(c)$,
simply by replacing the Fermi operators $a_{i\sigma}$ with the
transformed fermions $c_{i\sigma}$. Note that $O$ and $\tilde O$ are only
equivalent in the $U=\infty$ limit. The procedure which makes the Hubbard model
block diagonal is now well known\cite{eskes,spalek,sasha,girvin,anderson}.
One constructs a similarity
transformation $S$ which connects sectors that differ by at most one
`fictive' doubly occupied site such that
\beq
H=e^S\tilde H e^{-S}
\eeq
becomes block diagonal, where $\tilde H$ is expressed in terms of the transformed fermions. In the new basis,
$[H,\tilde V]=0$,
implying that double occupation of the transformed fermions
is a good quantum number, and all of the eigenstates
can be indexed as such. This does not mean that $[H,V]=0$. If it
were, there would have been no reason to do the similarity transformation in
the first place. $\tilde V$, and
not $V$, is conserved. Assuming that $V$ is the conserved
quantity results in a spurious local SU(2)\cite{lsu21,lsu22}
symmetry in the strong-coupling limit at
half-filling.

Our focus is on the relationship between the physical and `fictive'
fermions. To leading order\cite{eskes} in $t/U$, the bare fermions,
\beq
a_{i\sigma}&=&e^Sc_{i\sigma}e^{-S}
\simeq c_{i\sigma}-\frac{t}{U}\sum_{\langle j, i\rangle}
\left[(\tn_{j\bs}-\tn_{i\bs})c_{j\sigma}\right.\nonumber\\
&-&\left.c^\dagger_{j\bs}c_{i\sigma}c_{i\bs}+c^\dagger_{i\bs}c_{i\sigma}c_{j\bs}\right],
\eeq
are linear combinations of
multiparticle states in the transformed basis as is expected in
degenerate perturbation theory
By inverting this relationship, we find that to leading order, the transformed operator is simply,
\beq
c_{i\sigma}\simeq a_{i\sigma}+\frac{t}{U}\sum_{j} g_{ij}X_{ij\sigma}
\eeq
where
\beq
X_{ij\sigma}=\left[(n_{j\bs}-n_{i\bs})a_{j\sigma}-a^\dagger_{j\bs}a_{i\sigma}a_{i\bs}+a^\dagger_{i\bs}a_{i\sigma}a_{j\bs}\right].
\eeq
What we would like to know is what do the transformed fermions look
like in the lowest energy sector. We accomplish this by computing
the projected operator
\beq
(1-\tn_{i\bs})c_{i\sigma}&\simeq &(1-n_{i\bs})a_{i\sigma}+\frac{t}{U}\sum_{j}g_{ij}\left[
(1-n_{i\bs})X_{ij\sigma}\right.\nonumber\\
&&\left.-X^\dagger_{ij\bs}a_{i\bs}a_{i\sigma}-a^\dagger_{i\bs}X_{ij\bs}a_{i\sigma}\right].
\eeq
Simplifying, we find that
\beq\label{trans}
(1-\tn_{i\bs})c_{i\sigma}&\simeq &(1-n_{i\bs})a_{i\sigma}+\frac{t}{U}V_\sigma
a_{i\bs}^\dagger b_i\nonumber\\
&+&\frac{t}{U}\sum_{j}g_{ij}\left[
n_{j\bs}a_{j\sigma}+n_{i\bs}(1-n_{j\bs})a_{j\sigma}\right.\nonumber\\
&&\left.+(1-n_{j\bs})\left(a_{j\sigma}^\dagger
a_{i\sigma}-a_{j\sigma}a^\dagger_{i\sigma}\right)a_{i\bs}\right].
\eeq
Here $V_\sigma=-V_{\bar\sigma}=1$ and
$b_i= \sum_{j\sigma} V_\sigma c_{i\sigma}c_{j\bs}$ where $j$ is summed over the nearest neighbors of $i$.
As is evident, the projected `fictive' fermions involve the projected
bare fermion, $(1-n_{i\bs})a_{i\sigma}$, which yields the $2x$ sum
rule plus admixture with the doubly occupied sector
mediated by the $t/U$ corrections.  These $t/U$ terms, which are
entirely local and hence cannot be treated at the mean-field level, generate the $>2x$ or
the dynamical part of the spectral weight transfer.  This physics (which has been shown to play a significant role even at half-filling\cite{trem}) is absent from projected models such as the standard implementation\cite{lee,lee1,lee2} of the $t-J$ model
in which double occupancy is prohibited. As we have pointed out in the introduction, the physics left out by projecting out double occupancy is important because it tells us immediately that $L/n_h>1$; that is, new degrees of freedom must be present at low energy. Put another way, the operator in Eq. (\ref{trans}) is not a free excitation and but rather describes a non-Fermi liquid ($L/n_h>1$).   A process mediated by the
last term in Eq. (\ref{trans}), depicted in Fig. (\ref{fig1}), obtains only
if a doubly occupied and empty site are neighbors. This underscores
the fact that in Mott systems, holes can be heavily dressed by the upper
Hubbard band. It is this dressing that generates dynamical
spectral weight transfer.

Before we demonstrate how a single collective degree of freedom describes
such dressing, we focus on the low-energy Hamiltonian in the bare
electron basis. The answer in the transformed basis
is well-known\cite{eskes} and involves
the spin-exchange term as well as the three-site hopping term. Our
interest
is in what this model corresponds
to in terms of the bare electron operators which do not preserve
double occupancy. To accomplish this, we simply undo the
similarity transformation after we have projected the transformed
theory onto the lowest energy sector. Hence, the quantity of interest
is $H_{sc}=e^{-S}P_0e^SHe^{-S}P_0e^S$. Of course, without projection, the
answer in the original basis at each order of perturbation theory
would simply be the Hubbard
model. However, the question at hand is what does the low-energy
theory look like in the original electron basis. Answering this
question is independent of the high energy sectors in the transformed
basis because all such subspaces lie at least $U$ above the $m=0$
sector. Hence, it is sufficient to focus on $P_0e^SHe^{-S}P_0$. To
express $P_0e^SHe^{-S}P_0$ in the bare electron operators, we substitute
Eq. (\ref{trans}) into the first of Eqs. (14) of Eskes, et
al.\cite{eskes} to obtain
\begin{widetext}
\beq H_{sc}&=&e^{-S}P_0e^SHe^{-S}P_0e^S\nonumber\\
 &=& -t\sum_{\langle i,j\rangle}\xi_{i\sigma}^{\dagger}\xi_{j\sigma}-\frac{t^{2}}{U}\sum_{i}b_{i}^{(\xi)\dagger}b_{i}^{(\xi)}
-\frac{t^{2}}{U}\sum_{\langle i,j\rangle,\langle
i,k\rangle,\sigma}\left\{
\xi_{k\sigma}^{\dagger}\left[(1-n_{i\bar{\sigma}})\eta_{j\sigma}+\xi_{j\bar{\sigma}}^{\dagger}\xi_{i\bar{\sigma}}\eta_{i\sigma}+\xi_{i\bar{\sigma}}^{\dagger}\xi_{i\sigma}\eta_{j\bar{\sigma}}\right]+h.c.\right\}
\label{eq:sc1}
\eeq
\end{widetext}
as the low-energy theory in terms of the original electron
operators. Here, $\xi_{i\sigma}=a_{i\sigma}(1-n_{i\bar\sigma})$ and
$\eta_{i\sigma}=a_{i\sigma}n_{i\bar\sigma}$. The first two terms correspond to the $t-J$ model in
the bare electron basis plus 3-site hopping. However, terms in the bare-electron basis
which do not preserve the number of doubly
occupied sites explicitly appear. As
expected, the
matrix elements which connect sectors which differ by a single doubly
occupied site are reduced from the bare hopping $t$ to $t^2/U$. All
such terms arise from the fact that the transformed and bare electron
operators differ at finite $U$. Hence, Eq. (\ref{eq:sc1}) makes
transparent that the standard
implementation\cite{lee,lee1,lee2} of the $t-J$ model in which the transformed and bare
electron operators are assumed equal is inconsistent because the terms
which are dropped are precisely of the same order, namely $O(t^2/U)$,
as is the spin-exchange.

\section{New Approach}

Explicitly integrating out\cite{lowen1} the high-energy scale in the
Hubbard model uncloaks the collective degree of freedom that accounts
for the key difference between the bare and
transformed electrons and the ultimate origin of the breakdown of
Fermi liquid theory.
The central element of this theory is an elemental field, $D_i$, which we associated with the
creation of double occupation via a constraint. Our approach is in the spirit of Bohm and Pines\cite{bohm} who also extended the Hilbert space with a constrained field to decipher the collective behaviour of the interacting electron gas. A Lagrange multiplier
$\varphi_i$, the charge $2e$ bosonic field, enters the action much the way the
constraint $\sigma$ does in the non-linear $\sigma$ model. The
corresponding Euclidean Lagrangian is
\beq\label{LE}
{\cal L}&&=\int d^2\theta\left[\bar{\theta}\theta\sum_{i,\sigma}(1- n_{i,-\sigma}) a^\dagger_{i,\sigma}\dot a_{i,\sigma} +\sum_i D_i^\dagger\dot D_i\right.\nonumber\\
&&+U\sum_j D^\dagger_jD_j- t\sum_{i,j,\sigma}g_{ij}\left[ C_{ij,\sigma}a^\dagger_{i,\sigma}a_{j,\sigma}
+D_i^\dagger a^\dagger_{j,\sigma}a_{i,\sigma}D_j\right.\nonumber\\
&&+\left.\left.(D_j^\dagger \theta a_{i,\sigma}V_\sigma a_{j,-\sigma}+h.c.)\right]+H_{\rm con}\right]
\eeq
where
$C_{ij,\sigma}\equiv\bar\theta\theta\alpha_{ij,\sigma}\equiv\bar\theta\theta(1-n_{i,-\sigma})(1-n_{j,-\sigma})$
and $d^2\theta$ represents a complex Grassman integration. The constraint Hamiltonian $H_{\rm con}$ is taken to be
\beq\label{con}
H_{\rm con} = s\bar{\theta}\sum_j\varphi_j^\dagger (D_j-\theta a_{j,\uparrow}a_{j,\downarrow})+h.c.
\eeq
The Grassman variable $\theta$ is needed to
fermionize double occupancy so that it can properly be associated with the
high energy Fermi field, $D_i$. The constant $s$ has been inserted to
carry the units of energy. The precise value of $s$ will be determined
by comparing the low-energy transformed electron with that in
Eq. (\ref{trans}). This Lagrangian was constructed so that if we solve
the constraint, that is, integrate over $\varphi$ and then $D_i$, we
obtain exactly $\int d^2\theta \bar\theta\theta L_{\rm Hubb}=L_{\rm Hubb}$, the
Lagrangian of the Hubbard model.

The advantage of this construction,
however, is that we have been able to coarse-grain cleanly over the
physics on the scale $U$. That is, all the physics on the scale $U$ appears as the mass term of the new fermionic degree of freedom, $D_i$. It makes sense to integrate out $D_i$ as it
is a massive field in the new theory. The low-energy theory to
$O(t^2/U)$,
\beq
\label{HIR-simp}
H_{\rm eff}&=&-t\sum_{i,j,\sigma}g_{ij} \alpha_{ij\sigma}a^\dagger_{i,\sigma}a_{j,\sigma}\nonumber\\
&&-\frac{t^2}U \sum_{j} b^\dagger_{j} b_{j}-\frac{s^2}U\sum_{i}\varphi_i^\dagger \varphi_i\nonumber\\
&&-s\sum_j\varphi_j^\dagger a_{j,\uparrow}a_{j,\downarrow}-\frac{ts}U \sum_{i}\varphi^\dagger_i
b_{i}+h.c.\;\;.
\eeq
contains explicitly the charge $2e$ boson, $\varphi_i$. Here $b^{(a)}_i=\sum_{\sigma j}V_{\sigma}a_{i\sigma}a_{j\bar\sigma}$ where $j$ is the nearest-neighbour of $i$. To fix the
energy scale $s$, we determine how the electron operator transforms in
the exact theory. As is standard, we add a source term to the
starting Lagrangian which generates the canonical electron operator
when the constraint is solved. For hole-doping, the appropriate
transformation that yields the canonical electron operator in the UV
is
\beq
{\cal L}\rightarrow {\cal L}+\sum_{i,\sigma} J_{i,\sigma}\left[\bar\theta\theta(1-n_{i,-\sigma} ) a_{i,\sigma}^\dagger + V_\sigma D_i^\dagger \theta a_{i,-\sigma}\right] +
h.c.\nonumber
\eeq
However, in the IR in which we only integrate over the heavy degree of
freedom, $D_i$, the electron creation operator becomes
\beq\label{cop}
a^\dagger_{i,\sigma}&\rightarrow&(1-n_{i,-\sigma})a_{i,\sigma}^\dagger
+ V_\sigma \frac{t}{U} b_i a_{i,-\sigma}\nonumber\\
&+& V_\sigma \frac{s}{U}\varphi_i^\dagger a_{i,-\sigma}
\eeq
to linear order in $t/U$. This equation bares close resemblance to the
transformed electron operator in Eq. (\ref{trans}), as it should. In
fact, the first two terms are identical. The last term in
Eq. (\ref{trans}) is associated with double occupation. In
Eq. (\ref{cop}), this role is played by $\varphi_i$. Demanding that Eqs. (\ref{trans}) and (\ref{cop}) agree requires that $s= t$, thereby eliminating
any ambiguity associated with the constraint
field. Consequently, the complicated interactions appearing in
Eq. (\ref{eq:sc1}) as a result of the inequivalence between
$c_{i\sigma}$ and $a_{i\sigma}$ are replaced by a single charge $2e$ bosonic field
$\varphi_i$ which generates dynamical spectral weight transfer across the
Mott gap. The interaction in Fig. (\ref{fig1}), corresponding to the
second-order process in the term $\varphi_i^\dagger b_i$, is the key physical process that enters the dynamics at low-energy. That the dynamical spectral weight transfer can be captured
by a charge $2e$ bosonic degree of freedom is the key outcome of the
exact integration of the high-energy scale. This bosonic field represents a
collective excitation of the upper and lower Hubbard bands. However, we should not immediately conclude that $\varphi_i$ gives rise to a propagating charge $2e$ bosonic mode, as it does not have canonical kinetics; at the earliest, this could be generated at order $O(t^3/U^2)$ in perturbation theory. Alternatively, we believe that $\varphi$ appears as a bound degree of freedom. Since the
dominant process mediated by $\varphi_i$ requires a hole and a doubly
occupied site to be neighbours (see Fig. (\ref{fig1})) we identify $\varphi_i^\dagger a_{i\bar\sigma}$ as a new
charge $e$ excitation responsible for dynamical spectral weight
transfer. It is the appearance of this state at low-energy that
accounts for the breakdown of Fermi liquid theory in a doped Mott
insulator. Physically, $\varphi_i$ is the dressing of a hole by the
high-energy scale. We have
previously shown\cite{lowen1} that the formation of this bound state can produce the experimentally observed
bifurcation
of the electron dispersion below the chemical potential seen in
PbBi2212\cite{graf}, the mid-infrared band in the optical
conductivity and the pseudogap\cite{lowen2}.  Further, the breakup of
the bound state beyond a critical doping leads to $T-$ linear 
resistivity\cite{lowen2}.

The essential problem of Mottness is that in a hole-doped Mott
insulator, empty sites can arise from doping or from hopping processes
which mix the upper and lower Hubbard bands. Both contribute to $L$.
However, the spectral weight on the empty sites resulting from mixing
with the high-energy scale is proportional to $t/U$. Hence, such
empty sites effectively represent holes with fractional charge $-e(t/U)$ not 
$-e$ as is the the case with the holes resulting from doping. Consequently,
they make no contribution to $n_h$, thereby giving rise to $L/n_h>1$
for a doped Mott insulator and a general breakdown of the standard
Fermi liquid theory of metals.  At half-filling, such fractionally charged
sites still persist. Adding an electron to such a system at low
energies would require adding it coherently to a number of sites equal to $U/t\gg 1$. Such coherent addition of an electron at low energies has vanishing probability. The result is a gap for charge $e$ but not for
$e(t/U)$
excitations.  
At finite doping, holes in a Mott insulator are linear superpositions
of both kinds of empty sites. As a result, holes in the
hard-projected\cite{lee,lee1,lee2} $t-J$ model, in which $L=2x$, are not equivalent to holes in the Hubbard model. Approximations which prohibit explicit
double occupancy, such as the standard treatment\cite{lee,lee1,lee2} of the $t-J$ model in
which the operators are not transformed, miss completely\cite{lh1,prelovsek,haule} the localizing\cite{lh2,kotliar,choy} physics
resulting from the orthogonality between charge $e$ excitations and
the sites with spectral weight $t/U$. In the exact theory, the
physics associated with a finite length scale for double occupancy is
contained straightforwardly in a charge $2e$ bosonic field, instead of being buried
in complicated interaction terms in
Eq. (\ref{eq:sc1}).

As Polchinski\cite{polchinski}  (as well as others\cite{shankar}) have emphasized that from the point of
view of the renormalization group,  $T-$linear resistivity in the
cuprates makes a Fermi liquid description untenable.  We believe that
our low energy theory containing the charge $2e$ bosonic field is in
this sense a suitable replacement for Fermi liquid theory as it can
explain\cite{lowen2} $T-$linear resistivity.  
We have shown above that the bosonic field accounts for what would be
a consequence of complicated non-linear dependences on electron
operators in projected models. What is clear from
Polchinski's\cite{polchinski} arguments is that projected models do
not give a good basis upon which to build a theory -- they mask the
ubiquitous physics of strong coupling, namely that new degrees of freedom emerge at low energy.

\acknowledgements This work was initiated at the Kavli Institute for
Theoretical Physics and funded partially through PHY05-51164. We thank George Sawatzky for several conversations
that initiated this work, R. Bhatt, M. Hastings, R. Shankar, M. Sobol,
A. Chernyshev, A. -M. Tremblay, and O. Tchernyshyov for helpful discussions and the NSF, Grant Nos. DMR-0605769.


\begin{thebibliography}{99}
\bibitem{frac}In calculating $n_h$, one might be tempted to consider the total number of
empty sites, that is, those created by the dopants and those that
result from admixture
with the upper Hubbard band. Since the weight of the empty sites resulting from
admixing with the UV scale is not unity but $t/U$ to leading order, $n_h\approx 2x+2t/U=L>2x$. As a result, $n_h N_s$
is a non-integer, with $N_s$ the number of sites and as a consequence fractionally charge particles whose charge is determined by
the doping, the hybridization and the on-site repulsion are required. This is
unphysical. Strictly, for electron addition spectrum, $n_h N_s$ must
be an integer; consequently, $n_h=2x$.
\end{thebibliography}
\end{document}